\documentclass[12pt]{article}
\usepackage{amsmath}
\usepackage{graphicx,psfrag,epsf}
\usepackage{enumerate}
\usepackage{natbib}
\usepackage{url} 

\newcommand{\blind}{0}

\begin{document}

\def\spacingset#1{\renewcommand{\baselinestretch}%
{#1}\small\normalsize} \spacingset{1}


\if0\blind
{
  \title{\bf Challenges and Successes of Emergency Online Teaching in Statistics Courses
}
  \author{Analisa Flores \\
   Department of Statistics \\ University of California Riverside \\
   and \\
   Lauren Cappiello \\
   Department of Mathematics and Statistics \\ California State University Sacramento \\
   and \\
   Isaac Quintanilla Salinas \\
   Department of Mathematics \\ California State University Channel Islands
   }
  \maketitle
} \fi

\if1\blind
{
  \bigskip
  \bigskip
  \bigskip
  \begin{center}
    {\LARGE\bf Title}
\end{center}
  \medskip
} \fi

\bigskip
\begin{abstract}
As the COVID-19 pandemic took hold in early months of 2020, education at all levels was pushed to emergency fully remote, online formats. This emergency shift affected all aspects of teaching and learning with very little notice and often with limited resources. Educators were required to convert entire courses online and shift to remote instructional approaches practically overnight. Students found themselves enrolled in online courses without choice and struggling to adjust to their new learning environments. This article highlights some of the challenges and successes of teaching emergency online undergraduate statistics courses. In particular, we discuss challenges and successes related to (1) technology, (2) classroom community and feedback, and (3) student-content engagement. We also reflect on the opportunity to continue to enhance and enrich the learning experiences of our students by utilizing some of the lessons learned from emergency online teaching as new permanent online statistics courses are developed and/or moved back into the classroom.
\end{abstract}

\noindent%
\vfill

\newpage
\spacingset{1.45} 
\section{Introduction}
\label{sec:intro}

The COVID-19 pandemic has and continues to affect education at all levels for both instructors and students. Courses were quickly moved to an emergency, fully remote, online formats as the pandemic took hold in the early months of 2020. With little notice, no preparation, and limited resources, educators were required to convert entire courses online and shift to remote instructional approaches. Students were forced to enroll in online courses, scramble to obtain the necessary technology, and make decisions about housing, family responsibilities, and more. As a result, the development of emergency online courses by instructors and the learning environment experienced by students differed significantly from traditional online courses. 

In this paper, we present some of the challenges and successes of emergency online teaching specifically as they pertain to statistics and data science education. The three areas considered are 1) Technology, 2) Classroom Community and Feedback, and 3) Student-Content Engagement. We begin by summarizing some of the literature in these areas, followed by separate sections for each area of focus. Within each section, we discuss the most successful teaching strategies we found to mitigate these challenges. Finally, we consider ways to integrate these methods into the classroom as courses are moved back in person or offered in hybrid formats. Throughout each section, we consider the impacts on both students and instructors. Although many of the strategies found in this paper overlap with current findings in online education research, increased instructor flexibility remains the heart of the work done to create positive learning environments during the pandemic. 

\subsection{Technology}

The use of technology was essential to deliver material to students during the pandemic. With the unplanned migration to remote learning, educators faced several challenges in delivering mathematical and statistical concepts to students. \cite{patricia_aguilera-hermida_college_2020} and \cite{bower_technologymediated_2019} notes how students' lack of confidence in learning technology reduces their engagement in the material and lower students’ learning outcomes. Additionally, educators introducing new technology may have students focus on learning the technology instead of learning the course material \citep{lopez_pandemic_2021}. Lastly, educators noted that reviewing students' work on mathematical derivations posed a challenge virtually, especially if students did not have access to pens and tablets \citep{gilbert_everyone_2021}.

Contrary to the challenges, the use of technology brought new benefits to delivering course material \citep{robinson}. Several researchers noted the benefits of providing live-recorded lectures for their mathematics, statistics, and programming courses \citep{abrahamsson_comparison_2021, delgado_teaching_2021}; moreover, \cite{jones_engaging_2022} found the use of virtual whiteboards to be beneficial in delivering content to students. Educators found innovative ways to deliver their material to students; however, several educators agree that emergency remote learning takes a considerable amount to prepare material and does not have the same effectiveness as in-person instruction \citep{lopez_pandemic_2021}.

\subsection{Classroom Community and Feedback}

A sense of classroom community plays an important role in the student experience. Both sense of community and perceived student learning are strongly related to course satisfaction \citep{baturay}. However, students in science courses tend to feel a lower sense of community with both instructors and classmates as compared to students in non-science courses \citep{young}. A variety of strategies have been suggested to improve student sense of classroom community, including problem-based projects \citep{baturay2} and formative assessment \citep{gikandi}. Discussion boards may also increase a sense of classroom community, especially when posting is required and graded \citep{rovai}. Rovai also finds that stronger feelings of classroom connectedness relate to stronger feelings of satisfaction about educational goals (\citeyear{rovai}). 

Little has been written on providing effective feedback in the online classroom, especially with respect to handwritten mathematics and statistics assignments. However, feedback has been shown to promote the regulation of learning and is associated with both improved performance and higher levels of student satisfaction in the online learning environment \citep{espasa}.

\subsection{Student-Content Engagement}

Learner-content interaction is the way in which a student engages with course content to change their understanding or perspective \citep{moore}. Moore describes it as a defining characteristic of education. In a traditional face-to-face course, student-content engagement can be achieved by attending lectures, working on assignments, completing assessments, and more. In an online environment, student-content engagement can also include watching instructional videos and searching for information online \citep{abrami}. Findings show that the instructor's teaching style and visual presence are instrumental in engaging students with the content \citep{everett} and that both synchronous and asynchronous content delivery are effective methods for helping online students access content for critical student-content interaction \citep{banna}. Asynchronous formats specifically, provide students with more time to think critically and reflect on the material \citep{robinson}. 

A substantial amount of research exists that compares face-to-face and online strategies for engaging students with content. However, limited information is available on the effectiveness of strategies during emergency online learning; though it is emerging as a result of the COVID-19 pandemic. One such paper provides a discussion on effective engagement strategies, finding that students perceived student–content engagement strategies to be more effective than student–instructor and student–student strategies during emergency online learning \citep{abou}. Discipline specific strategies for online courses are sparse and nearly nonexistent for emergency online learning.

\subsection{Emergency Online Learning}

Despite significant literature on the online classroom, it is not immediately clear how emergency online learning impacts these findings. For example, much of the literature rests on the implicit assumption that students choose the online classroom. Are these ``best practices” the same for emergency remote classrooms in the midst of a global pandemic? How do we adapt pedagogy when students are forced into online learning or experiencing ongoing traumatic events? This paper addresses some of the ways in which we modified our pedagogy to support our students during the COVID-19 pandemic. We utilized existing approaches to online education where possible and made adjustments as necessary based on time constraints and student needs. Given the evolving situation with COVID-19 and many university administrations’ desire to see more online offerings, this is and will continue to be an important question in education research. 

\section{Technology}

As educators rushed to teach their courses online, many new challenges arose with delivering material virtually. Instructors and universities worked tirelessly to ensure students could continue their education remotely. This meant addressing challenges associated with the basic necessities of remote learning - ensuring access to things like computers, high-speed internet, video cameras, speakers, microphones, textbooks, and other required course materials. In addition, providing access to non-academic related services was of great concern (e.g., counseling and psychological services, health services, etc.). Furthermore, statistics educators experienced unique challenges in delivering courses due to the need for teaching statistical concepts, teaching programming methods, creating accessible material, and troubleshoot problems for students. 

\textbf{Use tablets and various note-taking software to present formulas, demonstrate problem solutions, and the ability to sketch visual aids.} Statistics programs have courses that are both applied and theoretical. Each case requires careful thought around delivering material. Elementary statistics courses introduce students to formulas for computing different probabilities and statistics. While instructors can provide formulas on slides, there is no replacement for hand-writing a solution to a problem and highlighting important concepts. Being able to manipulate a formula or draw a figure is necessary to explain ideas. To accommodate this need, we utilized tablets and various note-taking software. This allowed for the flexibility of easily manipulating a problem and providing visual aids to student questions. Afterwards, notes can be posted online for students to access and review as needed.

\textbf{Provide troubleshooting documentation and utilize accessible online platforms for statistical software.} Elementary statistics courses analyze data using statistical software. Regardless of the software, students need to gain access to or install the software on their computers. Detailed instructions were provided to install software and troubleshoot problems for students. To minimize the burden on students when installing software, a couple of approaches were deployed. We used online platforms such as RStudio Cloud (now called Posit Cloud\nocite{posit}) to teach R\nocite{rstats} without the concern of software or computing issues. Students on some campuses have access to virtual labs, which provide students with a virtual desktop filled with the applications they need to conduct analysis. Lastly, R packages were used to deliver material that students use to learn how to program in R. These packages can be updated to install other packages or data. For elementary statistics courses without a programming component, we took advantage of online resources such as the Rossman and Chance applet collection (\citeyear{rossmanchance}). The use of cloud platforms ensure that students have access to the computing resources needed for completing assignments.

In statistical computing courses, educators needed to find ways to deliver computational and programming concepts to students. Computing courses are easier to deliver online but developing material and having students work on both individual and group assignments were challenging. For many students, a statistical computing course requires a deep understanding of programming with statistical packages. Delivering programming material with current Learning Management Systems (LMS) poses new challenges. Additionally, debugging code via video conferencing or email is difficult since interacting with the computer to identify the error is limited. To address these issues, we utilized several tools designed by Posit for delivering content and troubleshooting code. We used RMarkdown\nocite{rmarkdown} documents to explain statistical computing concepts and provide  the corresponding R code. These documents can be converted to web pages that can be hosted on GitHub\nocite{github}. We also had students work with RMarkdown and send source files when debugging was necessary. This allowed us to collect the corresponding HTML file for assignments. For non-programming courses, Overleaf\nocite{overleaf} provided a nice alternative for Latex documents that allow students to write reports without worrying about the Tex installation process.

\textbf{Provide ‘practice’ assignments so students have the opportunity to get comfortable with new or unfamiliar technology.} In all types of statistics classes, homework and assessments were a significant challenge since we wanted to deliver accessible assignments virtually while maintaining academic integrity. For elementary statistics courses, instructors were able to provide some assignments and exams using online software or assessment tools from within their LMS. More advanced statistics courses that require students to show mathematical derivations of concepts called for alternative approaches such as having students submit scanned documents for homework and exams. This led to challenges with file formats and legibility. To ensure students learned how to scan documents with their smartphones, we had assignments at the beginning of the course with the sole purpose of letting students practice scanning their documents. We then provided feedback to notify students whose scanned documents were in an incorrect file format or were illegible. Accounting for students who have never used smartphone applications to scan documents or submit a file as a portable document file (PDF) is an important consideration. Allowing for practice and multiple attempts for submitting files saved time and frustration for both sides.

\section{Classroom Community and Feedback}

\textbf{Utilize discussion boards to build a sense of classroom community.} A strong sense of classroom community can be difficult to achieve in the online classroom. We were surprised, at first, to find that many students were not interested in speaking to one another in the Zoom classroom, even in small group breakout sessions\nocite{zoom}. Although students may have the opportunity to work together in a face-to-face statistics classroom, there are often few opportunities for deeper classroom discussion, especially in large lecture classes. The collaborative work accomplished in the classroom, such as think-pair-share activities or group work, is difficult to accomplish over Zoom and/or the LMS. Further, math anxiety may play a role in students’ hesitation to work together, even in face-to-face classrooms. As a result, the online statistics classroom may be more isolating than other disciplines. 

For the kind of student-student interaction needed for a strong community, we utilized discussion boards. It may be useful to create a student code of conduct for discussion boards and some of our colleagues used this as an opportunity to get students involved in creating their own community expectations. We found that including a student conduct statement in the course syllabus was a good way to set expectations for both discussion boards and the rest of the course. 

Discussion boards were graded based primarily on participation to encourage students to get involved in the classroom community in a low-pressure setting. Discussion board topics ranged from informal ``get to know you” type posts to posts designed to get students working collaboratively. Prompts might include asking students to share information such as their major and why they selected it, likes/dislikes of online learning, or extracurricular interests; academic strategies regarding time management or studying techniques; reflections on personal achievements or strengths/weaknesses and how to leverage them; or ‘just-for-fun’ prompts. For example, two recent just-for-fun prompts were ``If you could have any magical powers, what would you choose and how would you use them?” and ``Please share a fun fact (it doesn't have to be about you!)”. 

Students are also encouraged to respond to their peers and connect over shared majors, career goals, or hobbies. Students seemed to enjoy these discussion boards and prompts, expressing gratitude for attempting to engage with everyone in the class and fostering a sense of community instead of the ‘normal’ disconnected learning experience found in asynchronous courses. They appreciated seeing the instructor and teaching assistant posting answers to prompts, stating that it strengthens student-instructor connections. Instructors were also able to better understand their students’ perspectives, situations, and backgrounds, particularly in large courses.

Collaborative discussions related to course content might involve asking and answering questions (e.g., through an open Q\&A discussion board) or working on a challenging problem. One strategy is to assign a problem that most students will not be able to solve on their own and have them post an initial approach or solution. For example, the Monty Hall and Birthday problems are both solvable using concepts introduced in our introductory statistics courses, but tend to be challenging for students. As a class or in groups, they go through each other’s posts and are able to reach a solution collaboratively. This activity was initially suggested by a student, and several students mentioned they found it helpful to be able to see other students' approaches to problem-solving. However, it is important to note that in this setting, students often need reassurance upfront that they are not expected to get an answer right away and that you are more interested in the learning process than in ``correctness”.

\textbf{Conduct student check-ins, including asking for feedback regarding the course structure and what is and is not working well.} An additional consequence of remote learning was lack of interaction and nonverbal feedback between students and instructors, where instructors includes both lecturers and teaching assistants. Students often attended virtual class meetings with their cameras off, leaving the instructor to speak into the void. In person, presenters are in constant ``conversation” with the audience, monitoring verbal and nonverbal feedback. In a typical face-to-face conversation, people gain information on agreement and understanding from head and eye movements \citep{kleinke}. Even in a lecture-heavy classroom, these kinds of nonverbal cues provide conversational input which allow the instructor to adjust on the fly. In a virtual classroom where most students have their cameras off, this information is minimized or nonexistent. Further, when there are fewer communication cues available - as is the case in a virtual classroom with minimized point of view - the few available cues have a larger impact than they would otherwise \citep{walther1996}.

The lack of feedback often resulted in minimal direct student-teacher interaction. Less face-to-face time may make it more difficult for students to feel comfortable reaching out to faculty. As a result, we found that fewer students attended office hours. Further, students were unable to ask questions while interacting with asynchronous content. Compounding this problem was the difficulty in providing feedback while grading online. We did see an encouraging increase in student questions during live Zoom class sessions, possibly due to the ability to ask questions anonymously. However, these were difficult to monitor and students often struggled to ask mathematics questions over chat because of limitations related to mathematical notation. 

Check-ins can be conducted both privately (student-to-instructor) or on a course discussion board. Questions we found useful are provided here: (1) `What do you like about this class? What are some things that are working for you?' (2) `What is something that you enjoy from one of your other courses that we are not doing in this class?' (3) `What would your ideal course look like right now (given that we are online and at home)? Assume the instructor of your ideal course has unlimited time and resources.' (4) `How are you doing (both in this class and in general)? Are you holding up okay? Is there anything else you'd like to share with me? If you choose to write something, please let me know if it's okay for me to follow up with you about it.' Much of the student feedback referenced throughout this paper came from these check-ins. 

Students seemed to really appreciate the discussion board check-ins, commenting that it was nice to see how their classmates were feeling. In private responses to (4), multiple students noted that it meant a lot that someone had taken the time to ask how they were doing. On the other hand, many students chose to share how traumatic the pandemic had been for them, describing feelings of isolation, depression, and hopelessness. Checking in with students who are struggling can be a deeply painful experience and these responses were difficult to read and process. Instead, instructors may find it more manageable to recommend any student counseling services up front, making sure to mention that these services are free or low-cost for students. Students also seem receptive to hearing about their instructor’s use of these services when they were in school, or any other instructor experience that can help to demystify or destigmatize the process.

Another approach is to include a Student Support section in the course syllabus with information and links to campus support services such as Counseling and Psychological Services, Student Health Services, Food Pantry Services, Case Management Services, Title IX Office, and more. This might include brief descriptions of why and how a student would access these services. Having these resources easily accessible within the syllabus communicates to students that not only are we aware of the possibility that they may need to utilize them, but we also care about their well-being. Students may also appreciate your linking to these items on a separate LMS page, especially if your syllabus is fairly long. 

In the virtual classroom, it may be more difficult for students to feel comfortable connecting with instructors to ask for assistance. For further student-instructor interaction, we conducted check-ins on how students were doing academically. Each week, students answered a series of prompts including (1) `Write down one thing you learned that you have not thought about before', (2) `What questions or confusions do you have about this week’s content?' and (3) `Additional thoughts or comments about the material or the course'. Making a point to address questions - anonymously - in written materials or videos posted on the LMS helped to encourage students to reach out each week with their concerns. These types of weekly check-ins can also be done informally based on interactions with students via email or during office hours. Requesting questions outside of the classroom gives students an opportunity to think about or work with the material before asking for clarification. The questions then provide a scaffold for what topics may be reviewed or clarified each week. They also give the instructor insight into how to approach these topics in the future. 

\section{Student-Content Engagement}

Finding ways to engage students with course content can be a challenging endeavor regardless of the discipline and course format (face-to-face, hybrid, or remote).  This challenge was exacerbated in our online introductory level statistics courses where students were being exposed to unfamiliar terminology, concepts, and applications while being taught from a distance. Long Zoom\nocite{zoom} sessions alone are difficult to pay attention to; add in long lectures riddled with new terms, formulas, and ideas, and students quickly become overwhelmed. To combat these challenges and garner student engagement, we attempted various course structures in both synchronous and asynchronous formats with mixed success. 

\textbf{Provide student support and encourage engagement with course material through asynchronous content.} One of the simplest strategies is to record synchronous lectures and provide them to our students for review. This allowed students the flexibility to engage with course meetings outside of regular course time, as well as the ability to rewatch all or part of a discussion. We did not record office hours in order to give students time to freely ask questions regarding course material, as well as to ensure privacy for students needing to discuss other matters. 

To further promote flexibility and engagement, we shared asynchronous content such as short video ``micro-lessons”, informational pages, and external website links through the LMS. The primary component of the asynchronous material were micro-lessons. These short topic videos designed for asynchronous viewing were used to teach, review, or demonstrate concepts in a way that was both easy to find (without having to skip through an hour-long recording) and easy to pay attention to. The length of the recordings were 3-20 minutes, with a majority being less than 8 minutes. Keywords were added to video descriptions to make finding topics efficient and straightforward. The keywords intentionally consisted of foundational statistical terms (e.g., variability, mean, distribution, etc.) to help students connect ideas and review specific topics as needed. The recordings also provided an organized system for students to review material in preparation for quizzes and exams, or while working through homework assignments. These same micro-lessons have since been used during “the return to campus” both as complimentary material to face-to-face lectures and/or when instructors are absent due to conferences or illness. 

For instructors, creating the micro-lectures was extremely time consuming. In total, the authors of this article created over 400 short recordings.
However, the resulting benefit of the extensive reusability of the recordings, which can be incorporated into both future online and face-to-face learning environments makes the initial time commitment worthwhile. Another added benefit of the micro-lectures is the ability to create a semi-flipped classroom environment where students could watch the recordings ahead of time and the synchronous course meetings (whether online or face-to-face) could then be dedicated in part to running practice problems and answering student questions. This format introduces some new challenges associated with attendance issues and students’ dislike of utilizing breakout rooms for group work. We found that low attendance makes it difficult to utilize breakout rooms and when used, students either left the sessions (by signing out of Zoom) or were hesitant to interact with their classmates in these small group settings. 

Students appreciated the asynchronous content with many choosing to highlight their experience with the micro-lectures. In a Statistical Computing course, on average, 71\% of students watched the entire length of the video. For videos containing supplementary material, just 21\% of students watched the full length of the video - indicating that most students will only watch the required videos. Common themes from student feedback were that the short videos were easy to focus on; there was a clear delineation between topics; and videos could easily be paused and rewatched, watched with captions, or even watched at faster/slower speeds (particularly when studying for exams). Students also commented that it was helpful for them to watch the short content videos outside of class time and use the live time to ask questions and review the content through examples. Other supplementary materials provided to students were the use of websites and R Packages that contained code and explanations of statistical concepts. Written summaries and step-by-step examples were also provided. Students commented positively regarding websites that were accessible after the course for use in other classes.

\section{Future Considerations}

\subsection{Technology}

With emergency online learning, instructors were more willing to adopt new technology and deploy new strategies. As we have transitioned back to face-to-face classes, there are several technologies that we may continue to use in the classroom. The creation of videos and the use of online resources provides a repository of teaching materials that can be utilized in subsequent terms. Videos can be employed to create flipped classrooms, allowing class time to be used for discussions on statistical and programming concepts. Additionally, educators are incorporating new technology to manage their classrooms and deliver the material. Several instructors may use GitHub to collect assignments, host course material, and R packages. Instructors may continue using RMarkdown/Quarto\nocite{quarto} to create documents and presentations for their students.

As we move forward, there are several new technologies powered by artificial intelligence that will have an impact on how statistics courses will be taught. With the rise of ChatGPT \citep{openai}, students are able to ask questions and receive responses to both conceptual and programming questions. This presents new opportunities and challenges when teaching statistics and programming to students. While it can be a great tool to help students understand concepts or discover new programming techniques, it may also misguide students when they are not able to identify errors in the responses. We should caution students about using new technology such as ChatGPT as it may provide erroneous results. This technology will continue to improve over time and the Statistics community needs to begin discussing how it will shape the ways in which we educate our students. 

\subsection{Classroom Community and Feedback}

The modern student is used to socializing and finding information online. Emergency online teaching gave educators an opportunity to use this as a strength. To foster a sense of classroom community, it can be helpful to encourage community outside of the classroom itself. While in-person study groups can be useful for many students, virtual community spaces allow students to engage in more flexible ways. Instructors may wish to create discussion boards through the LMS or to allow students the freedom to create their own virtual spaces. Instructor-created discussion spaces can include opportunities for students to work together on the material or create connections. These may be especially helpful in large lectures, where students are often unable to interact with many of their peers. Additionally, students may create virtual spaces on other platforms such as Discord servers. Instructors may choose to interact in these spaces to answer student questions, or they may prefer to leave those spaces for students only. Either way, students should be expected to follow the institution’s code of conduct. A classroom code of conduct is also recommended and is an opportunity to get students involved in creating positive spaces.

To give feedback to students, the LMS allows the instructor to grade and provide feedback locally. However, the LMS is often poorly set up for grading mathematics and statistics work. Services like GradeScope\nocite{gradescope} are a more efficient way to provide feedback when grading online, especially for mathematics work. We recommend creating a practice assignment to get students used to submitting work on GradeScope/LMS which can help further streamline the process. 

To facilitate feedback from students, regular check-ins may be conducted through the LMS. These can be in the form of short weekly check-ins about the course material, or through mid-term surveys requesting feedback on the course. Weekly surveys can be used to keep track of student learning and areas where students are falling behind; mid-term surveys can be used to check in more broadly with how students feel about the class overall. These give the instructor an opportunity to review areas where students are falling behind and to address any student concerns. 

Finally, the classroom community may benefit from the incorporation of health and wellbeing. If time permits, instructors can request relevant offices to deliver presentations in their classrooms, or even take their students on field trips to offices and tutoring centers. Otherwise, instructors may find it helpful to reference these resources in several different places and throughout the term. For example, on the syllabus, through direct links in the LMS, and verbally before and after each important assessment.

\subsection{Student-Content Engagement}

By far, the biggest success for student engagement with course content is providing asynchronous material that can be accessed by students anytime, anywhere. Whether a course is taught fully online, hybrid, or face-to-face, the availability of course materials outside of the classroom has become a critical component of our course designs. For best use, the asynchronous material should be free and easily accessible. Some instructors may prefer creating their own material to ensure perfect alignment with their course organization and/or to have complete control over the content but should be forewarned that it is extremely time consuming and arduous when starting from scratch. Building a repository through each iteration of teaching a course is the best route so as not to become overwhelmed and to ensure high-quality material. Student creation of course content is also encouraged (such as creating a short topic video as a group term-project) but requires careful review for accuracy and permission should be granted from the students to use their videos as part of your course materials. 

An alternative to self-creating asynchronous course content, particularly for introductory level courses, is to access the ever-growing availability of online resources and open source textbooks that provide reading, problems, data sets, videos, and more. OpenIntro\nocite{openintro}, MyOpenMath\nocite{openmath}, and CourseKata\nocite{coursekata} are just a few examples of readily available and easily accessible resources for serving as primary or supplementary course material. For courses with a heavier focus on computation using R, a variety of free online textbooks can be found on the Bookdown\nocite{bookdown2} website with books ranging from R Programming to Data Visualization to Machine Learning and Regression. 

Finally, though students have the desire to have asynchronous material available, they may not  consistently utilize it. Supplementary or optional asynchronous material allows students to access additional resources as needed. However, if a course is designed so that the asynchronous material is mandatory, instructors should strongly consider assigning at least some portion of the course grade to it. Options that work well for grading access to asynchronous material include participation points, discussion posts, or quizzes (either embedded into the material or separately) which can be easily integrated into the LMS.

\section{Conclusion}

Through several terms and various course formats of emergency online teaching for a range of statistics courses, we consistently found the following strategies created an effective online learning environment for our students. 

\begin{enumerate}
    \item Use tablets and various note-taking software for presenting formulas, demonstrating problem solutions, and the ability to sketch visual aids.
    \item Provide troubleshooting documentation and utilize accessible online platforms for statistical software.
    \item Provide ‘practice’ assignments so students have the opportunity to get comfortable with new or unfamiliar technology. 
    \item Utilize discussion boards to build a sense of classroom community.
    \item Conduct student check-ins, including asking for feedback regarding the course structure.
    \item Provide student support and encourage engagement with course material through asynchronous content.
\end{enumerate}

\noindent As we have moved back to face-to-face, traditional online, and hybrid courses, these strategies have continued to be implemented with success. 

For future research, we are interested in understanding student and faculty perspectives with emergency online statistics courses at the beginning of the pandemic versus online courses as we are moving beyond the pandemic.

\section*{Acknowledgements}

We would like to thank the attendees of our Joint Statistical Meeting 2021 Birds of a Feather Discussion and our colleagues for sharing their own challenges and successes with us. We also appreciate the comments from the reviewers and editors at JSDSE. 

\bibliographystyle{apalike}
\bibliography{Bibliography-MM-MC}
\end{document}